\documentstyle[epsfig,12pt]{article}
\begin{document}
%
%
\newcommand{\x}{\cdot}
\newcommand{\ra}{\rightarrow}
\begin{titlepage}
\vspace{3 ex}
%
%
\vspace*{0.2truecm}
\begin{flushright}
\begin{tabular}{l}
CERN-TH/2002-248 \\
UAB-FT-531 \\
September 2002
\end{tabular}
\end{flushright}

\vspace*{1.0truecm}

\begin{center}
{
\LARGE \bf \rule{0mm}{7mm}{\boldmath  
Impact of B physics on model building\\
and vice versa: an example}\\
}

\vspace{4ex}

\vspace*{1.3truecm}

{\large
Joaquim Matias\\
}
\vspace{1 ex}

{\em
Theory Division, CERN, CH-1211 Geneve 23, Switzerland \\ and \\  
IFAE, Universitat Aut\`onoma de Barcelona, Spain \\
}
\vspace{2 ex}
%
%
\end{center}

\vspace{2 ex}
%
%

\vspace*{0.7truecm}

\begin{abstract}
We motivate that  the start-up of the $B$ factories has opened a 
new precision 
flavour physics era, with an important effect on model building. Using 
as 
an \, example a left--right model with
spontaneous CP violation, we will show how the inclusion 
of the new experimental data on $B$ physics
observables, together with the old observables coming from kaon physics,
has significantly widened our capacity to strongly constrain the 
parameter space up to the point
to exclude models.
On the contrary, using certain hypotheses, mainly
concerning isospin,  we  discuss how  theory 
may help us to
`test'  the data on charged, neutral and mixed
$B \to \pi K$ decays once experimental errors will be reduced.
\end{abstract}
  
\vspace*{1.3truecm}

\begin{center}
{\sl Invited talk at the\\
8th International Conference on
$B$ Physics at Hadron Machines -- BEAUTY2002\\
Santiago de Compostela, Spain, 17--21 June 2002\\
To appear in the Proceedings}
\end{center}

\end{titlepage}

%
\setlength{\oddsidemargin}{0 cm}
\setlength{\evensidemargin}{0 cm}
\setlength{\topmargin}{0.5 cm}
\setlength{\textheight}{22 cm}
\setlength{\textwidth}{16 cm}
\setcounter{totalnumber}{20}

\clearpage\mbox{}\clearpage

\pagestyle{plain}
\setcounter{page}{1}
%

In the post-LEP era, $B$ physics, following the example of the
fantastic accuracy achieved in the precision tests of the Standard
Model by the LEP experiments, could bring us the possibility to
open a new era of precision: the precision flavour physics era. A
huge experimental and theoretical effort will be necessary to
achieve this goal.

On the experimental side, the start-up of $B$ factories is
providing us with a cascade of new experimental data on $B$ meson
decays: $B_d \rightarrow J/\psi \; K_s$, $B \rightarrow \pi \pi$,
$B \rightarrow \pi K$, etc. Moreover, forthcoming hadronic
machines (LHCb~\cite{lhcb} and BTEV~\cite{btev}) will collect data on 
$B_s$ 
decay modes such as
$B_s \rightarrow J/\psi \phi$ and $B_s \rightarrow KK$, which
may open new avenues in the search for new physics.
On the theoretical side, we have two main tools to help us in this search:
CP violation in $B$ and $K$ physics and FCNC rare decays, both inclusive 
and 
exclusive.
Their study will offer precise and very valuable information on
new interactions associated to the flavour sector of the
fundamental theory that lies beyond the Standard Model.

$B$ physics provides us with a set of new CP-conserving and 
CP-violating observables in $B_d$ and $B_s$ meson decays (see
reviews \cite{gen,rob1}): $\Delta m_{Bq}$,
$\Delta \Gamma_q$, and sides and angles  
($\alpha$,$\beta$,$\gamma$) of the Unitarity Triangle (UT).
Some of these observables are related to the matrix element:
\begin{equation}\label{m12}
\langle M^0 | {\cal H}^{|\Delta| F=2}_{\rm eff}| \bar{M}^0\rangle =   
2 m_M \left(M_{12}^{\rm SM} + M_{12}^{\rm NP} + M_{12}^{\rm LD}\right)
\qquad F=S ({\rm kaons}), B ({ B\; {\rm mesons}})  
\end{equation}
where $m_M$ stands for the corresponding kaon or $B$ meson mass;
$M_{12}^{SM}$ is the SM contribution, $M_{12}^{\rm NP}$ 
the new physics contribution and $M_{12}^{\rm LD}$  the $(\Delta F=1)^2$
contributions. Mass differences in the $B_{d,s}$ system are obtained from 
(\ref{m12}):
\begin{equation}
\Delta m_{Bq}=2 \left| M_{12}^{(q)} \right| \qquad \qquad q=d,s.
\end{equation}

In order to disentangle new physics effects in the mass difference 
\cite{alilondon}, it 
is important to recall that this observable is afflicted by hadronic 
uncertainties coming from $f_{Bq}^2 B_q$. Concerning the weak mixing 
phase
\begin{equation}
\phi^{B_q}_M ={\rm arg}\,M_{12}^{B_q},
\end{equation}
it measures the angle $2 \beta$ in the $B_d$ system and a very small angle 
$\delta \gamma$ in the $B_s$ system. The sine of $2 \beta$  can be 
obtained from the CP asymmetry in $B^0_d\to J/\psi K^0_S$ \cite{cart} or 
the sides of 
the UT, while the cosine, in particular its sign, is a very interesting 
future 
observable. The corresponding angle in the $B_s$ case is negligibly small 
in the SM, i.e. the corresponding CP asymmetries are small, so it is an 
excellent place to look for new physics.

Regarding the other two angles, $\gamma$ can be 
obtained from 
non-leptonic B decays 
such as 
$B\to \pi K$ \cite{all}--\cite{quim} and the sides of the 
UT 
and, finally, $\alpha$ 
it is 
traditionally analysed using isospin in $B \to \pi \pi$ \cite{gronlon} or 
using SU(3) plus dynamical assumptions and factorization \cite{GRppn}, 
although an 
interesting alternative \cite{our2} to  control 
the hadronic 
penguin parameters is  
to use the decay modes $B_d \to \pi \pi$ and $B_s \to KK$ (or $B_d \to 
\pi K$) to extract, instead, $\gamma$
and together with $\beta$  obtain $\alpha$ \cite{our1,our2,rfx}.

The width difference 
$\Delta \Gamma_q$
in the $B_d$
system is expected to be too small to be measurable, but it could be
non-negligible for the $B_s$ meson. Unfortunately, in the presence of new 
physics it 
can   
only decrease \cite{grossman}. 

The second tool in the search for new physics is the analysis of rare $B$ 
meson decays 
(see \cite{isi} for a review). These are processes that are suppressed at 
tree 
level, i.e. new physics  can compete  on the same
footing as the SM. They allow us to put constraints on the
parameter space of models beyond the SM. One of the more important 
rare decay is 
the 
inclusive $B\to X_s \gamma$ that provides information on the magnitude of 
the Wilson coefficient $C_7$ and it has been evaluated very precisely in 
the SM~\cite{misiak} and 
supersymmetry~\cite{gg}\footnote{There is also an exclusive, very rare 
mode 
related 
to this:
$B_s \to \gamma \gamma$. It has also been evaluated in the SM 
\cite{bsggsm}
and 
supersymmetry \cite{bsgg}.}. 
Other important rare modes are the inclusive and exclusive semileptonic 
decay modes driven by the 
quark 
transition $b \to s l^ + l^-$, whose forward--backward 
asymmetry~\cite{agm}\footnote{Important progress with a full NLO 
calculation of this
asymmetry has been reported in \cite{bfs}.}
 provides 
information on the sign of $C_7$ and also on $C_9$ and 
$C_{10}$.



Here we will focus mainly on the first tool. 
In sections 1 to 3, we will show an example of the impact that new 
experimental data 
coming from $B$ physics are having on 
model
building. We will see how a specific type of 
left--right model with spontaneous CP violation~\cite{fbm} gets strongly 
constrained and 
could even be 
excluded, thanks to the new observables coming from $B$
physics and their combined analysis with the old  $K$ physics observables.
In section 4 we will try to argue the other way around and we will  
use  theory together with 
certain reasonable assumptions to `test' data  on $B \to \pi K$ decays.

\section{Description of the model: left-right model with
spontaneous CP}

This model is based on the gauge group $SU(2)_L\times SU(2)_R
\times U(1)$ with the feature that CP violation is spontaneous 
originating from a phase in the vacuum expectation values. 
There is a lot of  literature on left--right models \cite{fbm}--\cite{lrn} 
with and 
without spontaneous CP violation.

The Spontaneously Broken Left--Right Model (SB--LR)  has the interesting 
properties of being  fully testable and
distinct from the SM. The particle content of this model,
concerning the quark, gauge and scalar sectors, consist of left and
right quark doublets:
$$
q_{Li} = \left( \begin{array}{c} U_i\\D_i\\ \end{array}\right)_L
\sim (2,1,1/6)
\qquad q_{Ri} = \left( \begin{array}{c} U_i\\D_i\\
 \end{array}\right)_R
\sim (1,2,1/6)
$$
that acquire their masses via a spontaneous breakdown of the
symmetry such that the bidoublet $\phi$ acquires a vev

$$
\Phi = \left( \begin{array}[c]{c@{\:\:}c}
 \phi_0^1 & \phi_1^+\\ \phi_2^-
&
    \phi_0^2 \end{array} \right)\sim (2,\overline{2},0) \quad
\rightarrow  \quad \; \; \langle \Phi \rangle = {1
\over{\sqrt{2}}} \left(\begin{array}{c@{\:\:}c} v & 0\\ 0 &
w\end{array}\right).
$$
In order to complete the breakdown of the symmetry group to
$U(1)_{em}$ respecting the LR symmetry, two other triplets (or
doublets) are required:
$$
\chi_L = \left( \begin{array}{c} \chi_L^{++}\\ \chi_L^+\\
\chi_L^0
\end{array}
\right) \sim (3,1,2), \quad \chi_R = \left( \begin{array}{c}
  \chi_R^{++} \\ \chi_R^+\\ \chi_R^0\end{array} \right) \sim
(1,3,2) \quad \rightarrow \quad \langle \chi_{L,R}\rangle = {1
\over{\sqrt{2}}}\left(
\begin{array}{c}0\\ 0\\v_{L,R} \end{array}
\right).
$$

The scalar sector then contains a SM-like neutral scalar, a single
charged scalar, 
and two neutral scalars, with flavour-changing
couplings to quarks.

The relevant parameters for the rest of the discussion coming from
the scalar sector will be the ratio of vev of the bidoublet
$r=|v/w|$ and the phase  $\alpha=\arg(vw)$. However, we will use  a more
convenient combination of them called $\beta$ and $\beta^ 
\prime$  \footnote{These combinations are useful to calculate 
the phases of the CKM matrices,  because the dependence on 
$\beta^\prime$ becomes trivial. Moreover, for a natural choice of 
parameters 
$\beta^\prime$  
is negligibly small.}  defined 
by \cite{frere1}: $$\beta 
= \arctan \frac{2 r \sin\alpha}{1-r^2} \qquad e^{i \beta^\prime}={1
- r^2 e^{-2i\alpha} \over |1-r^2e^{-2i\alpha}|}.
$$

Other parameters are the  mass of the charged and flavour-changing 
scalars. However, since
they are both taken to be heavy, the neutral FC scalars cannot mix
with the light scalar and must be nearly degenerate \cite{Ecker:vv}, we 
take a common mass parameter for the charged and neutral FC
scalars $M_H$.

Concerning the gauge sector, in addition to the usual charged
$W_L^\pm$ and neutral $Z_L$ gauge bosons we have an extra charged
$W_R^\pm$ and neutral $Z_R$ gauge boson. Only the charged gauge
bosons will be relevant to our discussion. The spontaneous
breakdown of the symmetry generates also the mass of the 
charged gauge bosons:
$$
M_{W^\pm}^2 = \left( \begin{array}{cc} \displaystyle
\frac{g_L^2}{4}\,
    ( 2 v_L^2 + |v|^2 + |w|^2) & -g_L g_R v^* w /2 \\
-g_L g_R v w^* /2 & \displaystyle \frac{g_R^2}{4}\, (2 v_R^2 +
|v|^2 + |w|^2)\end{array} \right) \equiv \left( \begin{array}{cc}
M_L^2 & M_{LR}^2\,e^{-i\lambda}
    \\ M_{LR}^2\,e^{i\lambda} & M_R^2\end{array}\right).
$$
The mixing between the two physical charged $W$ bosons is
$$
\left( \begin{array}{c} W_1^+\\ W_2^+\end{array}\right) = \left[
\begin{array}{cc} \cos\zeta & -e^{i\lambda} \sin\zeta \\
e^{-i\lambda}\sin\zeta &
  \cos\zeta\end{array} \right] \left( \begin{array}{c} W_L^+\\
W_R^+\end{array}\right)
$$
where the $W_L$--$W_R$ mixing angle is defined as
$$
\tan 2\zeta = -\frac{2M_{LR}^2}{M_R^2-M_L^2}.\, 
$$
The charged current reads (with $g\equiv g_L\equiv g_R$ and
without displaying unphysical scalars and charged Higgs
contributions):
\begin{eqnarray*}
{\cal L}_{cc} & = & -\frac{g}{\sqrt{2}}\, \bar{U}_i \left[
\cos\zeta
  (V_L)_{ij} \gamma^\mu P_L - e^{-i\lambda} \sin \zeta (V_R)_{ij}
  \gamma^\mu P_R \right] D_j\, W^+_{1\mu}\\
& & {}-\frac{g}{\sqrt{2}}\, \bar{U}_i \left[ e^{i\lambda}\sin\zeta
  (V_L)_{ij} \gamma^\mu P_L + \cos \zeta (V_R)_{ij}
  \gamma^\mu P_R \right] D_j\, W^+_{2\mu},
\end{eqnarray*}
giving rise to two CKM matrices one left ($V_L$) and one right
($V_R$). The phase structure of these matrices will be explained
in the next section.

\section{Quark mixing, phases and parameters}

The combination of P invariance together with spontaneous CP violation
imposes strong restrictions on the coupling matrices of the Yukawa 
interaction part of the lagrangian,

\begin{equation}\label{eq:real}
-{\cal L}_Y = \Gamma_{ij} \bar q_{Li} \Phi q_{Rj} + \Delta_{ij} \bar
q_{Li}
 \widetilde{\Phi} q_{Rj} + {\rm h.c.}
\end{equation}

Both coupling matrices $\Delta$ and $\Gamma$ are taken to be real and 
symmetric\footnote{There are special cases where this is not so, which 
are 
not discussed here.}. This is crucial, because it means that we can 
diagonalize the mass matrices by only two unitary matrices 
\begin{equation}\label{eq:UV}
M^{(u)} = U D^{(u)} U^T,\quad M^{(d)} = V D^{(d)} V^T,
\end{equation}
where  $D^{(u,d)}$ are diagonal mass matrices. We have in this model two 
CKM matrices; in the basis where the coupling matrices are 
symmetric, these are related to one another by the following relation:
$$
K \equiv K_L = U^\dagger V = K_R^*.
$$
Those models not fulfilling this constraint are not affected by the 
present 
analysis \cite{lrn}.
These CKM matrices can be written in a more standard form:
$K_L$ in standard CKM form with a unique phase $\delta$ and
 $K_R$ containing the remaining  5 new phases: $\alpha_1$, 
$\alpha_2$,
$\alpha_3$,
$\epsilon_1$, $\epsilon_2$.

The important point to recall  is that all phases (including $\delta$) can 
be 
expressed as an exact function of $m_{u,d}$, $r$,
$\alpha$, $V_{ij}$.

Finally,  we collect here all  new 
parameters of the model that we have introduced up to now:
\begin{itemize}
\item $ M_2 \sim O(1\,{\rm TeV})$,  mass of
right-handed gauge boson;
\item $ \zeta$, the mixing angle between $W_R$ and $W_L$,
$\zeta\geq 0$;
\item $ g_R$, the coupling of $W_R$. We set here $g_R=g_L$.
\item $ 0\leq r\leq 1$ and $ 0\leq \alpha\leq \pi$
(or better $\beta$), parametrizing
the spontaneous breakdown of
CP. We will indeed work in the so-called ``natural'' region for  $r\sim 
O(m_b/m_t)\sim 0.02$ that explains the observed smallness of the CKM 
mixing angles. This value implies $\zeta \sim O(10^{-4})$.
\item degenerate extra Higgs masses $ M_H \sim
O(10\,{\rm
TeV})$, and we assume $M_H > M_2$.
\item quark { mass signs}, $2^5=32$-fold ambiguity (mass
ratios). Moreover, we distinguish two values of $ \delta$ in case of no
CP violation: $ \delta=0$ (CLASS I solutions) and
$ \delta=\pi$
(CLASS II solutions). Notice that the sign of the quark masses is
an observable 
in SB--LR models.
\end{itemize}

\section{Impact of SB--LR on observables}

Before starting the discussion of the constraints that the model receives
from $B$ and $K$ physics observables, it is important to discuss the 
behaviour of the model in two general 
regimes
 for the common Higgs mass 
($M_H$) and the extra charged gauge boson ($M_2$):

\begin{itemize}
\item
Decoupling limit: this corresponds to $M_2,M_H\rightarrow\infty$. In this 
limit we observe that the CKM phase $\delta$ gets strongly restricted
 $|\delta^{SBLR}|<0.25$ for class I solutions and 
$|\delta^{SBLR}-\pi|<0.25$ for class II. However, a global fit 
\cite{parodi} 
yields $\delta=1.0\pm0.2$. This implies that the SM limit of this model
 is inconsistent by $3.5\sigma$ with current experiments.

\item Finite $M_2$ and $M_H$ masses: the gauge boson mass $M_2$ entering 
through 
the mixing angle $\zeta$ or as a propagator cannot induce observable 
effects at tree level. However, at loop level, sizeable effects are 
expected, because the Inami--Lin functions in a LR model are larger 
roughly by a 
factor of 4  with respect to the SM and the suppression due to the mixing 
angle
is compensated by large quark mass terms from spin-flips $\zeta 
\rightarrow \zeta
m_t/m_b$ inside the loop 
in $b \rightarrow s \gamma$. Concerning the Higgs, their contribution is 
also heavily suppressed at tree level by factors  $(m_W/M_H)^2\sim 
10^{-4}$. We have the contributions of neutral FC Higgs to $\Delta 
F=2$ processes at tree level and the charged Higgs contributions enhanced 
by $m_t/m_b$ in b penguins. Finally,  the gauge and Higgs 
contributions are similar in size in $K$ physics observables, while the 
Higgs 
contribution dominates 
in $B$ mixing.   
\end{itemize}

\subsection{Constraints from the $K$ system}

The observables that we considered are $\Delta M_K$, $
\epsilon_K$ and
$\epsilon^\prime$. They are
plagued by
theoretical uncertainties.
For example, the long distance contributions $M_{12}^{\rm LD}$ 
(\ref{m12}) in the $K$ system are not 
very 
well known, but
they are expected to be sizeable. Then one is forced to make a reasonable 
assumption 
concerning  $\Delta m_K$: the LR contribution should at most saturate 
$\Delta m_K$, i.e. 
$2 |M_{12}^{\rm K,LR}| <
\Delta
m_K^{\rm  exp}$.
The second observable, $\epsilon$,  provides us with information 
about the 
phase difference between $M_{12}$ and $\Gamma_{12}$. In order to check 
the usual formula used for $\epsilon$ we rederive it from 
\begin{equation}
\label{epsilon2} \epsilon = \frac{1}{2\sqrt{2}}\, e^{i\pi/4}\sin\left(
\arg
M_{12} + 2
  \arg a_0\right),\end{equation}
avoiding approximations in 
the SM that need not  be valid in the 
SB--LR. Following all the procedure \cite{fbm} one can show that: a) phase 
redefinitions of $V_L$ and $V_R$ cancel in 
(\ref{epsilon2}), b) from the experimental result $ |\epsilon|^{\rm 
exp} = (2.280\pm
0.013)\times
10^{-3}$, the following bound can be obtained: $\arg M_{12} + 2
\arg a_0
= (6.449 \pm   
0.037)\times  10^{-3}$. 
Some remarks are in order here. First, since both terms are small and also 
${\rm
  Re}\,M_{12}\approx |M_{12}|$ (\ref{epsilon2}) can be reduced to the 
standard formula for $\epsilon$. Second, while in the SM  $\arg a_0$ is 
neglected,
in LR-SB this is no longer true and the $W_R$ 
contribution computed in \cite{Ecker:vv} is: 
$ 2 |\arg a_0^{\rm LR}|
<
0.005\cdot \left(\frac{1\,{\rm TeV}}{M_2}\right)^2$. However, given the 
uncertainties involved in the computation of this contribution and the 
lack of Higgs contribution, we prefer to include  this term twice as an 
uncertainty of $\arg M_{12}$ (assuming the Higgs contribution to be 
smaller than the extra charged gauge boson):
\begin{eqnarray}
\label{uiui}
\hspace*{-.7cm}6.375\times 10^{-3}-0.01  \left(\frac{1\,{\rm
TeV}}{M_2}\right)^2<  &\widetilde{\theta}_{M}& < 6.523\times
10^{-3} +
0.01  \left(\frac{1\,{\rm TeV}}{M_2}\right)^2 
\\
{\rm  with\ } &\widetilde{\theta}_{M}& =
\left|\frac{2\, {\rm Re}\,M_{12}}{\Delta m_K}\right| \arg
M_{12}.\nonumber
\end{eqnarray}

Finally, concerning $\epsilon^\prime$, we have been extremely conservative 
and  have only required the SB--LR model to predict correctly the sign of 
this observable.

\begin{figure}[t]
$$
\epsffile{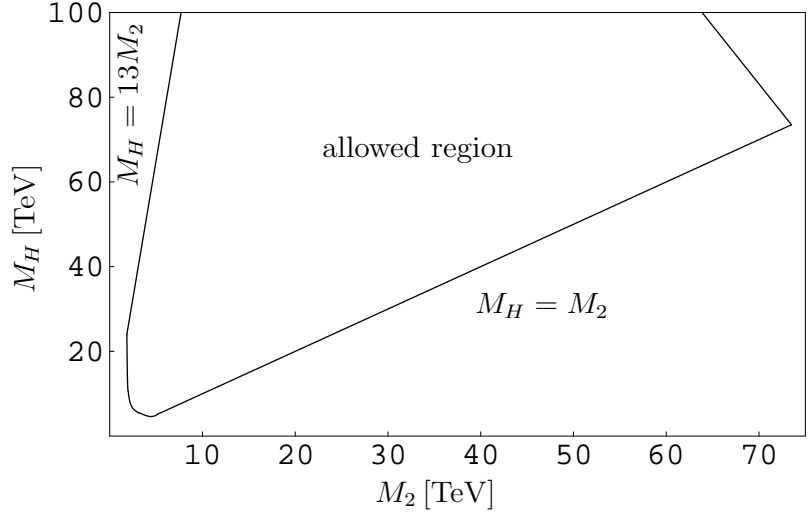}
$$
\caption[]{Allowed values for $M_2$ and $M_H$ from the $K$ physics 
constraints.
}\label{fig:kfig}
\end{figure}

The consequence  of imposing the 
previous constraints 
on the parameter space of this model
are mainly of two types \cite{fbm}:

\begin{itemize}

\item[a)] As expected from the previous discussion, in the decoupling 
limit 
$M_2,M_H\to\infty$, we obtain from the constraint on $\epsilon$ 
(\ref{uiui}):
$$
\widetilde{\theta}_{M} < 2.9\times 10^{-3},
$$
which implies again that this limit is excluded by the smallness of 
the CKM phase $\delta$ in the 
SB--LR model.

\item[b)] In the finite mass case we get lower and upper bounds:

\begin{itemize}

\item   
{ Lower bounds}  on the extra boson masses:
\begin{equation}\label{eq:idiotie1}
M_2 > 1.85\,{\rm TeV},\qquad M_H > 5.2\,{\rm TeV}. \nonumber
\end{equation}

The bound on $M_2$  is the usual one, while the bound on $M_H$ is 
lower, since we included not only the  charm contribution but also the 
top
quark contribution (destructive interference), which was  usually 
neglected.

\item
{ Upper bounds}  on the extra boson masses, not very constraining:
\begin{equation}\label{eq:idiotie2}
M_2 < 73.5\, {\rm TeV},\qquad M_H < 230\,{\rm TeV}. \nonumber
\end{equation}
\end{itemize}
\end{itemize}
These results are illustrated in Fig.~\ref{fig:kfig}.
Interestingly, once these bounds are combined with those coming from $B$ 
physics, the allowed region of parameter space of this model gets strongly 
reduced, as we will show in the following sections.

\subsection{$B$ physics constraints}

In this section, we  discuss the constraints coming from 
$B^0$--$\bar{ B}^0$ Mixing. First those coming from the mass 
difference
$\Delta m_{B_d}$ and $\Delta m_{B_s}$, and then the CP asymmetry 
$B^0_d\to J/\psi K^0_S$.

\subsubsection{$\Delta m_{B_d}$ and $\Delta m_{B_s}$}

\begin{figure}[t]
$$
\hspace*{-.8cm}{\epsfclipon
        \psfig{file=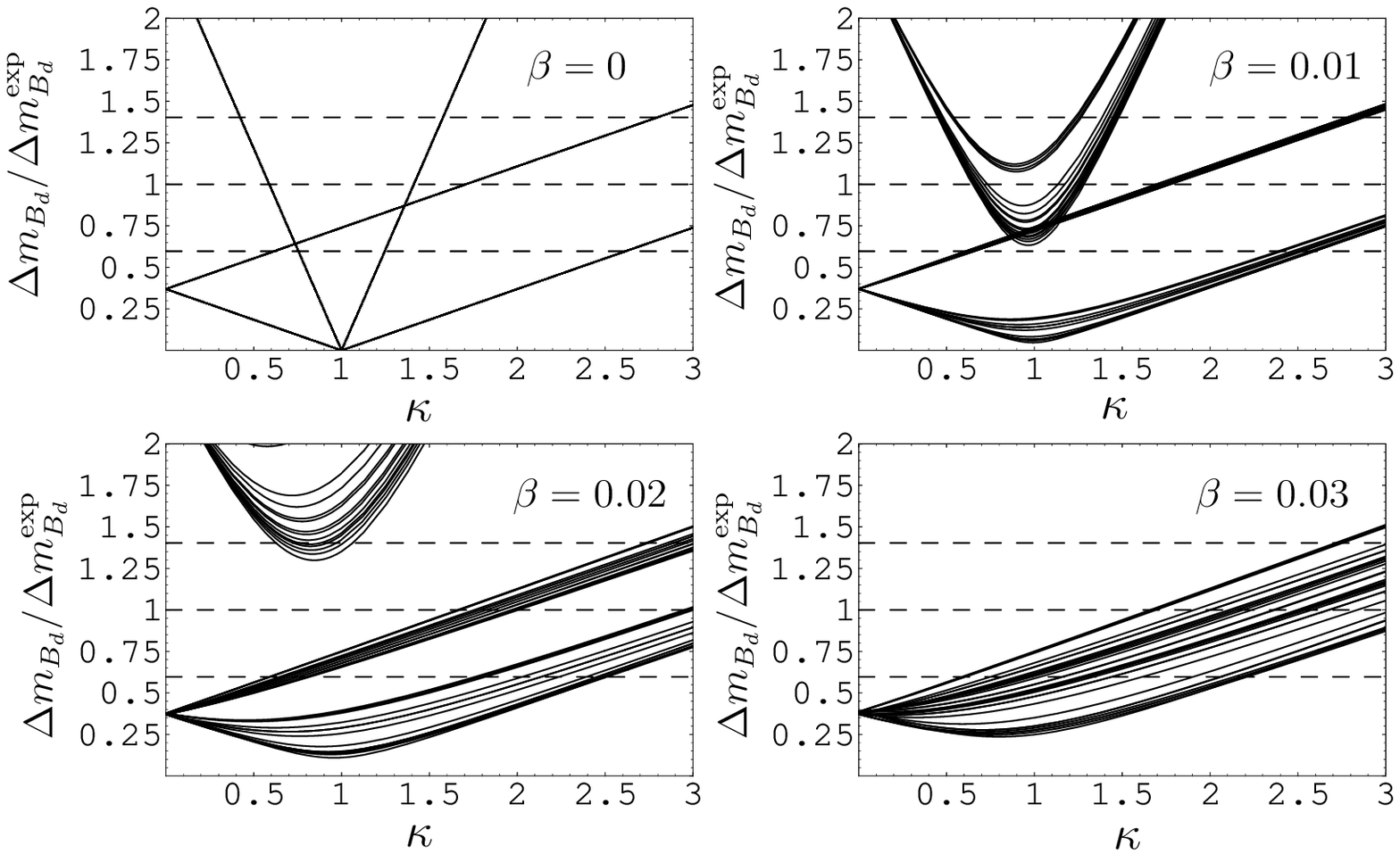,bb=0 -20 236 143 , width=6.5cm}}
\; \; {\epsfclipon
        \psfig{file=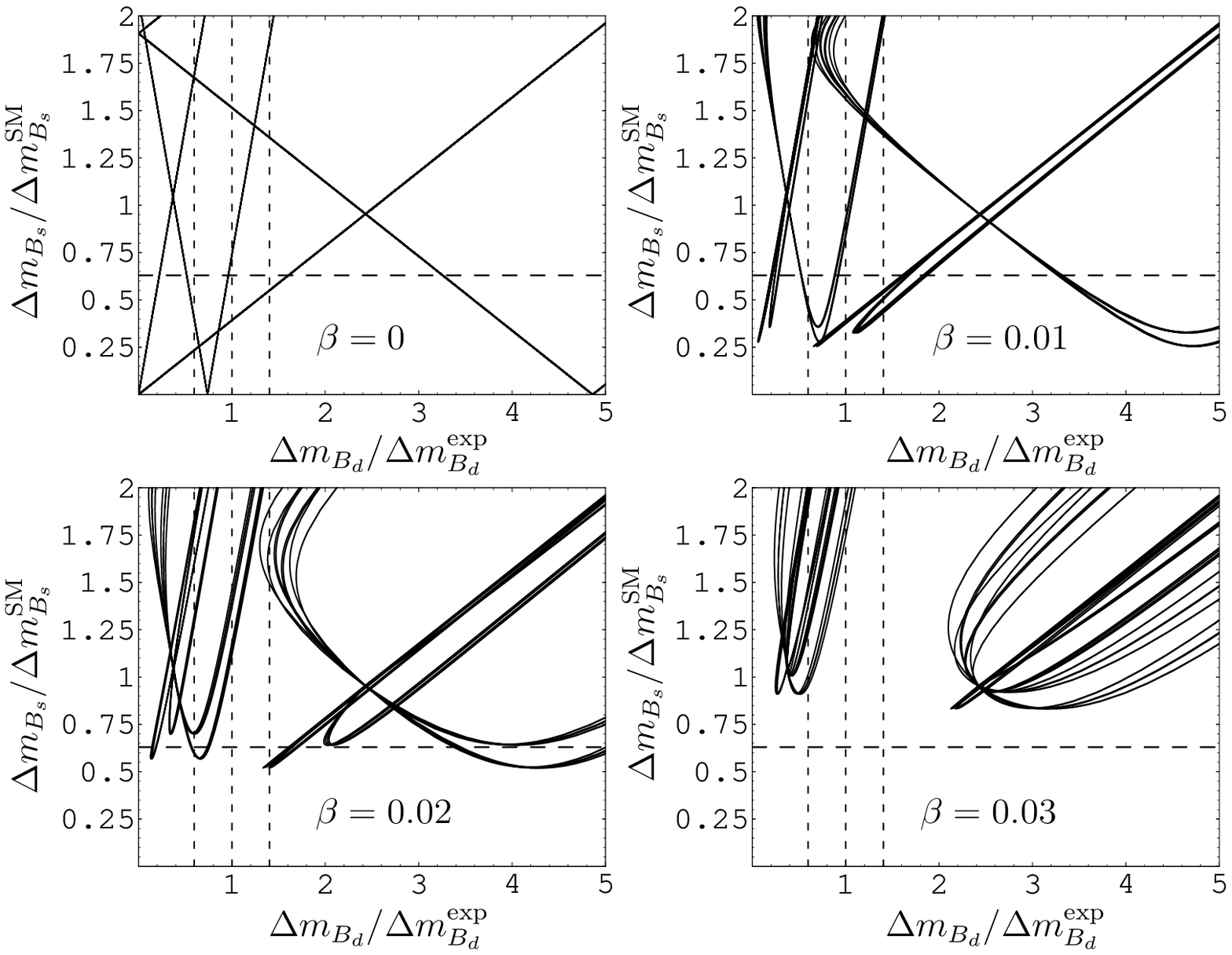,bb=235 0 476 180 ,
width=6.1cm}}
$$
\vspace*{-0.6cm}
\caption[]{First plot: Constraints from $\Delta m_{B_d}$. 
Dashed
lines are the experimental result and theory errors.
The lower curves are class I solutions, the upper curves are
class II. Second plot: Correlation between $\Delta m_{B_d}$ and $\Delta 
m_{B_s}$. Short dashes denote experimental results and theory error for 
$\Delta m_{B_d}$, long dashes  the lower bound on $\Delta m_{B_s}$. Left 
lines are class I and right ones class II.}
\label{plot12} 
\end{figure}

In the SB--LR model, $M_{12}$ gets 3 types of new contributions:
\begin{equation} 
M_{12}  =  M_{12}^{\rm SM} + M_{12}^{W_1W_2} + M_{12}^{S_1W_2}
+
M_{12}^H \equiv  M_{12}^{\rm SM} + M_{12}^{\rm LR}.\nonumber
\end{equation} 
The new contributions are box diagrams including $W_R$ and
unphysical 
scalars, and tree-level neutral Higgs exchanges.
If we write the total $M_{12}$ in a more compact form \cite{fbm}
\begin{eqnarray}
M_{12} & = & M_{12}^{\rm SM} ( 1 + \kappa\, e^{i\sigma_q} ), \nonumber \\
{\rm with\ }\kappa & \equiv & \left| \frac{M_{12}^{\rm LR}}{M_{12}^{\rm
      SM}}\right|, \nonumber\\
\sigma_q & \equiv & {\rm arg}\, \frac{M_{12}^{\rm LR}}{M_{12}^{\rm
    SM}}\ =\ {\rm arg} \left( -\frac{V_{tb}^R V_{tq}^{R*}}{V_{tb}^L
V_{tq}^{L*}} \right).\label{eq:def_sigma}
\end{eqnarray}
we find numerically for the quasi-spectator independent $\kappa$,
\begin{equation}\label{eq:kappa}
\kappa = \frac{B_B^S(m_b)}{B_B(m_b)} \left[ \left(\frac{7\,{\rm
        TeV}}{M_H}\right)^2 + \eta_2^{LR}(m_b) \left( \frac{1.6\,{\rm
        TeV}}{M_2} \right)^2 \left\{0.051 - 0.013 \ln
\left(\frac{1.6\,{\rm
        TeV}}{M_2} \right)^2\right\}\right],
\end{equation}
where $\eta_2^{LR}(m_b) \approx 1.7$ is a LO short-distance correction 
and the ratio of bag factors evaluated from QCD sum rules or to leading 
order in $1/N_{c}$ is ${B_B^S(m_b)/B_B(m_b)} = 1.2 \pm 0.2$.
The constraint on $\Delta m_{B_d}$ and the bound on 
$\Delta m_{B_s}$ translate into a 
constraint on 
$\kappa$.
From the experimental results
\begin{equation}\label{eq:delmBd}
\Delta m_{B_d} = (0.472\pm 0.016)\,{\rm ps}^{-1} \qquad {\rm and} \qquad 
\Delta 
m_{B_s} > 
12.4\,{\rm ps}^{-1},
\end{equation}
and from the expression for  $\Delta m_{B_d}$, we obtain
\begin{equation}\label{eq:5.25}
\left| (V_{tb}^L V_{td}^{L*})^2 (1+\kappa \, e^{i\sigma_d}) \right| =
  (6.7\pm 2.7)\times 10^{-5},
\end{equation}
which translates into 
$$
\kappa < 3,
$$
while the lower bound on $\Delta m_{B_s}$ implies
\begin{equation}\label{eq:boundmBs}
\left|(V_{tb}^L V_{ts}^{L*})^2 (1 + \kappa e^{i\sigma_s})\right| > 9.6
\times 10^{-4}.
\end{equation}
However, since the ratio between the mass differences seems to have a 
smaller theoretical error, owing to lattice results on this particular 
combination of hadronic parameters, it is interesting to use also the 
following bound

$$
\left| \frac{(V_{tb}^L
    V_{ts}^{L*})^2 (1+\kappa e^{i\sigma_s})}{(V_{tb}^L
    V_{td}^{L*})^2 (1+\kappa e^{i\sigma_d})}\right|=
\frac{\Delta m_{B_s}}{\Delta m_{B_d}}
\frac{m_{B_d}}{m_{B_s}} \left( \frac{f_{B_d}}{f_{B_s}}\right)^2
\frac{\hat{B}_{B_d}}{\hat{B}_{B_s}}
 > 17.2.
$$ 
The main implications  of these constraints, illustrated in 
Fig.~\ref{plot12}, are that an upper bound on 
$\kappa$ is obtained, that the decoupling limit that corresponds here to 
($\kappa 
\rightarrow 0$) is excluded, as can be seen in the first 
plot of Fig.~\ref{plot12}. 
Also, class 
II solutions are excluded for $\beta \geq 0.021$ (the upper curves of 
the first plot of Fig.~\ref{plot12} go 
further up for these values of $\beta$) and, finally, class I solutions 
require $\kappa>0.52$ and class II $\kappa>0.42$, which are mainly driven 
by the Higgs contribution that become essential. The SB--LR can naturally 
accommodate any value of $\Delta m_{B_s}$ larger than the SM value, as 
illustrated in the second plot of Fig.~\ref{plot12}.

\subsubsection{CP asymmetry
$B^0_d\to J/\psi K^0_S$}

The gold observable at present coming from $B$ physics is the 
time-dependent CP asymmetry in $B^0_d\to J/\psi K^0_S$, the 
experimental averaged 
number for this asymmetry is \cite{expsinwa}:
\begin{equation}\label{eq:CDF} 
a_{\rm CP} = \frac{\Gamma(\bar B_d^0(t)\to J/\psi K_S^0) -
\Gamma(B_d^0(t)\to J/\psi K_S^0)}{\Gamma(B_d^0(t)\to J/\psi
K_S^0) +
\Gamma(\bar B_d^0(t)\to J/\psi K_S^0)} =
(0.78\pm0.08)\,
\sin\,(\Delta m_B t). \nonumber   
\end{equation}
where we assume vanishing direct CP violation. This observable can be 
written 
$$ 
a_{\rm CP} = {\rm Im}\, \lambda \; \sin (\Delta m_B t),
$$
In the SM, ${\rm Im}\, \lambda$ measures directly the angle $\beta$ of the 
Unitarity Triangle
\begin{equation}\label{eq:betaCKM} 
{\rm Im}\, \lambda=\sin  2\beta_{\rm CKM} \quad \quad 
\beta_{\rm CKM} =
\arg \left( -\frac{V_{cd}^L V_{cb}^{L*}}{V_{td}^L
V_{tb}^{L*}}\right), \nonumber
\end{equation} 
while in the SB--LR symmetric model this observable measures
\begin{eqnarray} 
\hspace*{-.9cm}{\rm Im}\, \lambda(B^0\to J/\psi K_S^0) 
&\!\!=\!\!&\sin  2\beta^{\rm eff}_{\rm CKM} \nonumber \\
&\!\!=\!\!&\sin \left[ 2
\beta_{\rm CKM}  +
\arg \left( 1 + \kappa e^{i \sigma_d}\right) -
 \arg \left( 1 + \frac{M_{12}^{\rm K,LR}}{M_{12}^{\rm
K,SM}}\right)\right]. \nonumber
\end{eqnarray}   
\begin{figure}[t]
\vspace*{-1cm}
$$
{\epsfclipon
        \psfig{file=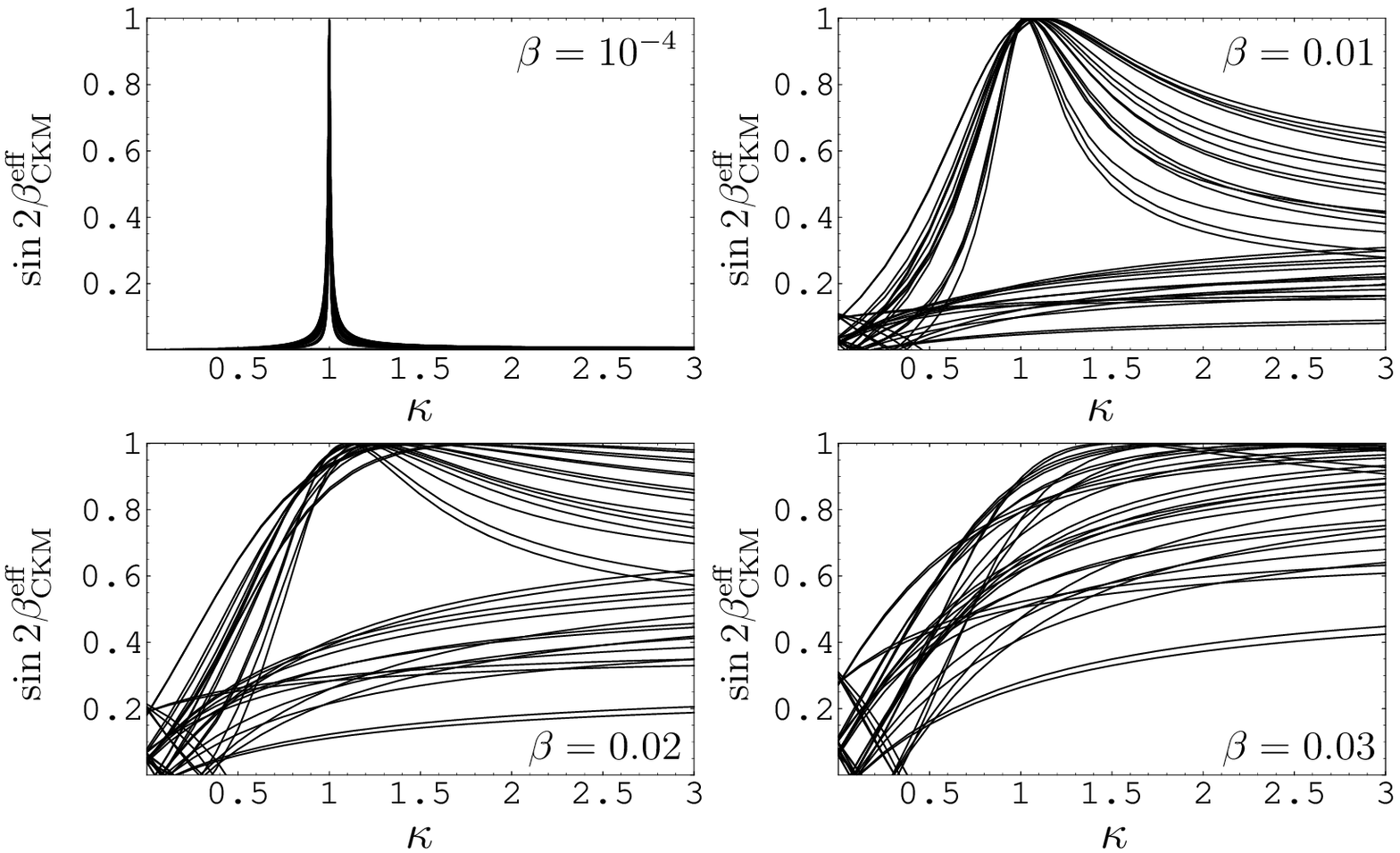,bb=238 146 474 291 ,
width=6.8cm}} {\epsfclipon 
        \psfig{file=sin2beta.eps,bb=0 0 236 145 ,
width=6.8cm}}
$$ \vspace*{-30pt}
\caption[]{$\sin 2\beta^{\rm eff}_{\rm CKM}$ in the SB--LR as a function 
of
  $\kappa$ for several values of $\beta$.
} \label{jiji}
\end{figure}
The constraints obtained from this observable are illustrated in 
Fig.~\ref{jiji}: all negative  values for $\sin 2\beta$
are excluded, the SM expectation for  $\sin 2\beta^{\rm
eff}_{\rm CKM}\approx 0.75$ can be easily
accommodated, for $\beta < 0.03$ there are two branches, one with small 
$\sin
2\beta^{\rm eff}_{\rm CKM}<0.4$, the other with all possible values 
between 0 and 1. Finally, if we impose  $\sin 2\beta^{\rm
eff}_{\rm CKM}$ to be around its
SM expectation this implies $\kappa \approx 0.6$ or
$\kappa>1.2$.
These are the independent constraints that the CP asymmetry in $B^0_d\to 
J/\psi K^0_S$ imposes in the parameter space. The final step will be to 
combine all of them.

\subsection{Combining constraints: results and consequences}

Up to this point we have seen, after considering  the 
observables $\Delta m_K$, $\Delta m_{B_d}$,
$\Delta m_{B_s}$, $\epsilon$, $\epsilon'$ (only sign) and $\sin 
2\beta_{\rm
  CKM}^{\rm eff}$, that a SB--LR model based on the gauge group 
$SU(2)_L\times SU(2)_R \times U(1)$ with spontaneous CP violation 
reproduces easily the 
experimental results for CP-conserving observables.

But, once we combine CP-violating observables from the old $K$ physics 
observables with the new $B$ physics observables, they become very
restrictive: a strong anticorrelation is found between the signs of 
Re$\,\epsilon$ and $\sin 2\beta_{\rm CKM}^{\rm eff}$.

The combination of all these constraints yields the following 
results\cite{fbm}:

\begin{itemize}
\item all but one quark mass signatures are { excluded}; 
\item predictions for $\Delta m_{B_s}$ are in the range
  $ (0.6-1.1)\,\Delta m_{B_s}^{\rm SM,exp}$, i.e.  not much information 
expected from this observable if it falls in the SM range;
\item
the { parameter space} for gauge and scalar masses are
strongly
restricted when combining both $K$ and $B$ constraints.
$$ 2.75 \; {\rm TeV}<M_{W_{R}}<13, \; {\rm TeV} \quad
10.2 \; {\rm TeV}<M_{H}<14.6 \; {\rm TeV}.
$$   

\end{itemize}

Finally, as the main conclusions,  the SM limit of the SB--LR model 
is excluded by more than $4 \sigma$ and  the maximal value of 
$a_{\rm 
CP}(B\to J/\psi K_S)\equiv
\sin 2
\beta_{\rm
CKM}^{\rm
    eff} < 0.1$ because of the anticorrelation mentioned above, 
is 
incompatible with the present experimental
result by several $\sigma$, pointing to a possible ruling out of this 
model
in its present form.

\begin{figure}[t]
{\epsfysize=0.35\textwidth\epsffile{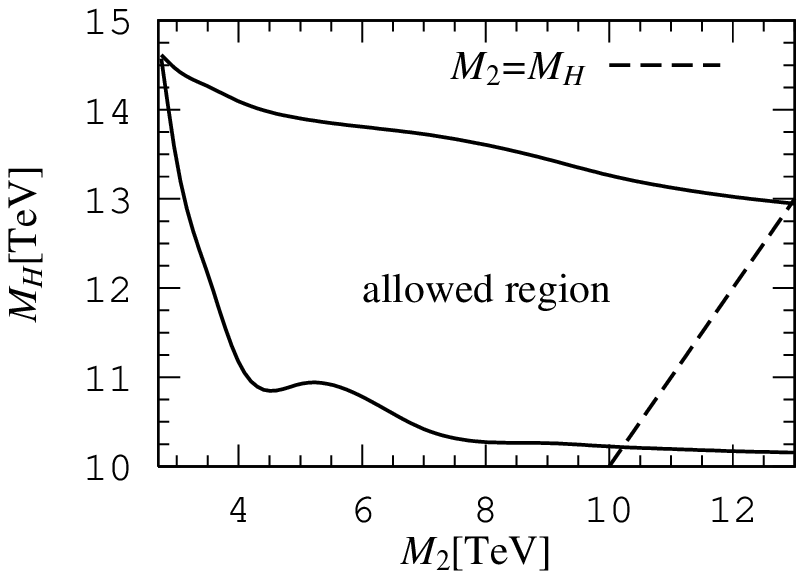}}  
{\epsfysize=0.35\textwidth\epsffile{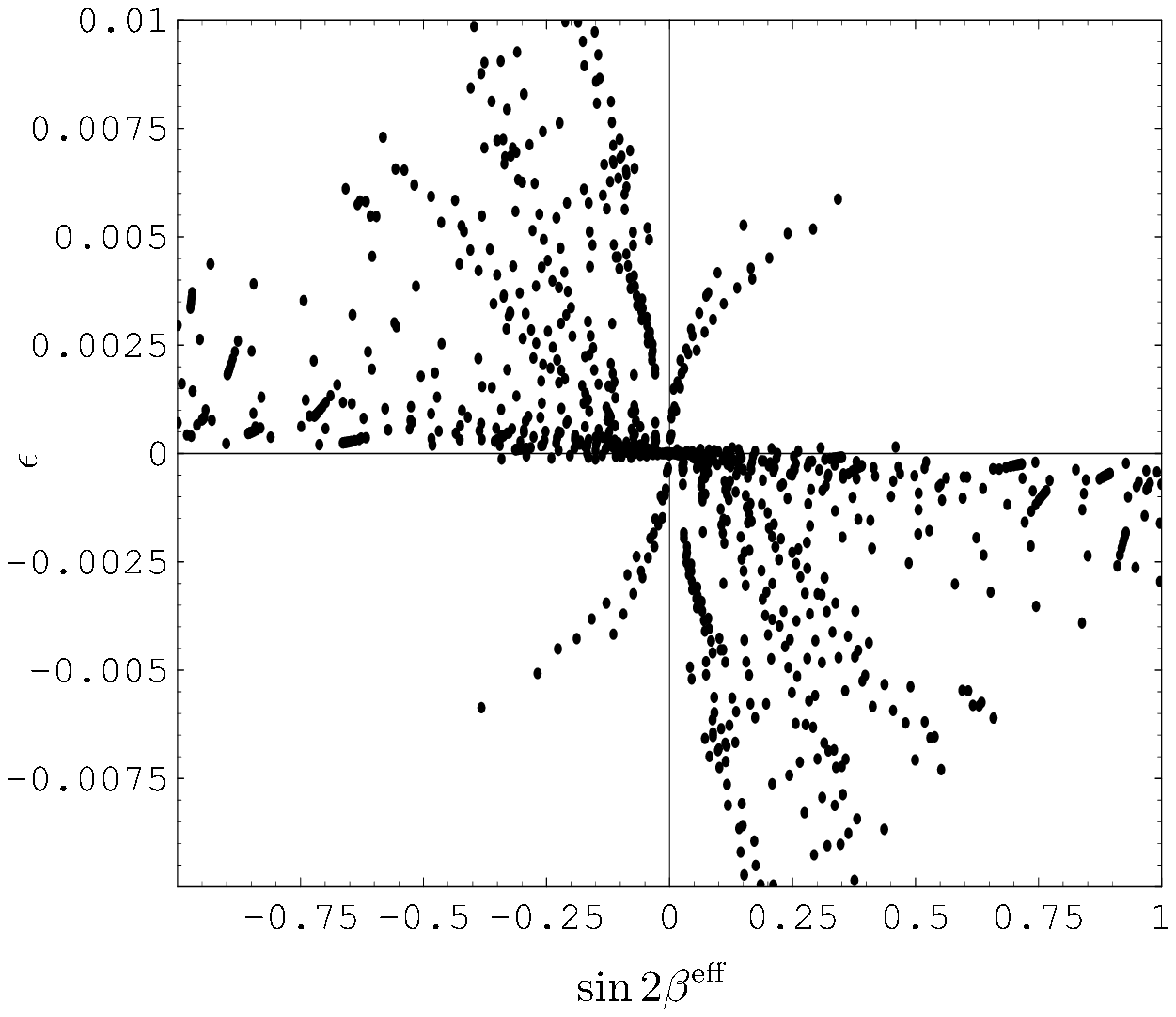}}
\caption[]{First plot: Allowed region in $(M_2,M_H)$, taking into account 
all
constraints. Second plot: Allowed values for the CP-violating parameters 
$\epsilon$  
  and $\sin 2\beta_{\rm CKM}^{\rm eff}$ after imposing the other 
constraints.
} 
\end{figure}

\section{Vice versa: model independent sum rules to `test' data on
$B\to \pi K$}

Here, we will try to argue the other way around, and
show how theory together with some reasonable hypothesis can help us, in a 
model-independent way, to `test',
in a certain sense, data on $B\to \pi K$ decays. We will construct a set 
of relations or sum
rules \cite{n1,quim,lip,gronaur} relating different observables of $B\to 
\pi 
K$ decays.

Rare $B$ decays based on the quark transition  $\bar b\to\bar s
q\bar q$ are described by the effective hamiltonian \cite{buchalla}:
\begin{displaymath}
{\cal H}_{\rm eff} = \frac{G_F}{\sqrt{2}} \left\{ \sum_{i=1,2} C_i
\left(\lambda_u Q_i^u+\lambda_c Q_i^c \right) - \lambda_t
\sum_{i=3}^{10} C_i Q_i \right\}+{\rm h.c.},
\end{displaymath}
{where $Q_{1,2}^{c}$ are current--current operators  and $Q_{3-6}$
are QCD-penguin operators; both induce a change of isospin
($\Delta I=0$), while $Q_{1,2}^{u}$ are current--current operators
and $Q_{7-10}$ are electroweak operators generating a change of
isospin ($\Delta I=0,1$). }

Since isospin plays a fundamental role in this discussion, it is
helpful to recall a few points. An initial $B$ meson state
$|B\rangle$ has isospin {$I=1/2$} and a final state $| \pi K
\rangle$ can have isospin {$I=1/2, 3/2$}. As a consequence, the
amplitude of a $B \rightarrow \pi K$ process can be decomposed in
three pieces:
\begin{eqnarray} { {\cal A}(B \rightarrow \pi K)}={ D_{\frac{1}{2} \; 0}}+{
A_{\frac{1}{2} \; 1}}+ { A_{\frac{3}{2} \; 1}},  \end{eqnarray}
where the subindices ${(D,A)}_{ I_{\pi K} \; \Delta I}$ stand for
${ I_{\pi K}}$ final state of isospin and ${ \Delta I}$
  change of isospin of the b quark transition.
$D$ refers to the dominant contribution coming from QCD penguins,
and $A_{\frac{1}{2} \; 1},A_{\frac{3}{2} \; 1}$ contain 
contributions  from electroweak
penguins and current--current operators but not QCD penguins. New physics 
contributes in particular to $A_{\frac{1}{2} \; 1}$ and $ A_{\frac{3}{2} 
\; 
1}$, and 
they, therefore, are the interesting pieces to measure.

$B \rightarrow  \pi K$ decays are described using CP-averaged
branching ratios \cite{fl2},\cite{n1}:
\begin{eqnarray} \label{eqa}
R_{\;}&=&{
 } \left[{{\rm BR}(B_{d}^0
\rightarrow \pi^- K^+) + {\rm BR}({\overline{ B_{d}^0}}
\rightarrow \pi^+ K^- ) \over {\rm BR}(B^+ \rightarrow \pi^+ K^0)
+ {\rm BR}(B^- \rightarrow \pi^- {\overline{ K^0} } )} \right],
\nonumber
 \\ \label{eqa2}
R_{c}&=&{2} \left[{ {\rm BR}(B^+ \rightarrow \pi^0 K^+) + {\rm
BR}(B^- \rightarrow \pi^0 {K^-} ) \over {\rm BR}(B^+ \rightarrow
\pi^+ K^0) + {\rm BR}(B^- \rightarrow \pi^- {\overline{ K^0}} ) }
\right], \nonumber
 \\ \label{eqa3}
R_{0}&=&{2 {
 }} \left[{{\rm BR}(B_{d}^0
\rightarrow \pi^0 K^0) + {\rm BR}(\overline{B_{d}^0} \rightarrow
\pi^0  \overline{K^0} ) \over {\rm BR}(B^+  \rightarrow \pi^+ K^0)
+ {\rm BR}(B^- \rightarrow \pi^- \overline{K^0} )} \right].
\end{eqnarray}
We will use these definitions here, because, in those terms,
 the expressions for the sum rules become  simpler.
Other definitions for the charged and neutral channels that are
used in the literature are $R_{*}=1/R_{c}$ \cite{n1} and
$R_{n}=R/R_{0}$ \cite{fl2}.
 CP asymmetries are the second
type of observables:
\begin{eqnarray} \label{eqabis}
{\cal A}_{\rm CP}^{0 +}&=&{{\rm BR}( B^+ \rightarrow \pi^0 K^+ )-
{\rm BR}( B^- \rightarrow \pi^0 { K^-} ) \over {\rm BR}( B^+
\rightarrow \pi^0 K^+ )+ {\rm BR}( B^- \rightarrow \pi^0 { K^- }
)},
\nonumber \\
{\cal A}_{\rm CP}^{+ 0}&=&{{\rm BR}( B^+ \rightarrow \pi^+ K^0 )-
{\rm BR}( B^- \rightarrow \pi^- \overline{ K^0} ) \over {\rm BR}(
B^+ \rightarrow \pi^+ K^0 )+ {\rm BR}( B^- \rightarrow \pi^-
\overline{K^0 } )},
\nonumber \\
{\cal A}_{\rm CP}^{- +}&=&{{\rm BR}( B_{d}^0 \rightarrow \pi^- K^+
)- {\rm BR}( {\overline{ B_{d}^{0}}} \rightarrow \pi^+ K^- ) \over
{\rm BR}( B_{d}^0 \rightarrow \pi^- K^+ )+ {\rm BR} ( \overline{
B_{d}^0} \rightarrow \pi^+ K^- )},
\nonumber \\
{\cal A}_{\rm CP}^{0 0}&=&{{\rm BR}( B_{d}^0 \rightarrow \pi^0 K^0
)- {\rm BR}(\overline{ B_{d}^0} \rightarrow \pi^0 \overline{ K^0}
) \over {\rm BR}( B_{d}^0 \rightarrow \pi^0 K^0 )+ {\rm BR}(
\overline{B_{d}^0} \rightarrow \pi^0 \overline{ K^0 } )}.
\end{eqnarray}


Generically, the CP-averaged branching ratios can be written in
the form $1+ \alpha{ A_{\frac{1}{2} \; 1}}+\beta {
A_{\frac{3}{2} \; 1}}$, where $\alpha$ and $\beta$ are constants
that depend on the particular CP-averaged branching ratio that we
are describing. Therefore, we can attach a physical meaning to
these observables $R$, $R_c$, $R_0$, as a measure of the physics
that violates isospin, either standard or new. In other words, if
there was no isospin violation, $A_{\frac{1}{2} \;
1}=A_{\frac{3}{2} \; 1}=0$ and all CP-averaged branching ratios
would measure 1.
Consequently, we can write these observables in the following
form \cite{quim}:
\begin{eqnarray} \label{ere1}
R&=&1+u_{+},\nonumber
\\ \label{ere2}
R_{c}&=&1+ z_{+},\nonumber
\\ \label{ere3}
R_0 &=&1 + n_+.
\end{eqnarray}
Using isospin decomposition we can show that
\begin{eqnarray}
\label{equuz1} u_+ &\sim& {\cal O} (r) + {\cal O} (r^2, r\rho, q_C
r) \nonumber  \\
 z_+ &\sim& {\cal O} (r_c) + {\cal O} (r_c^2,
r_c\rho, q r_c).
\end{eqnarray}
If we assign a generic value $\epsilon$ ($<1$) to the small 
parameters
$r, r_c$  and $q,q_c,\rho$ then $u_+$
and $z_+$ are quantities of ${\cal O}(\epsilon)$ in this sense.
Moreover, it is possible to relate $n_+$ with
the other two parameters $u_+$ and $z_+$ up to a quantity of
${\cal O}(\epsilon^2)$:
\begin{equation}\label{equnnn} n_+ = u_+ - z_+ + k_1 \quad \quad {\rm
with} \quad  k_1 \sim {\cal O} (r_c r,...) \sim {\cal
O}(\epsilon^2). \end{equation}
Therefore, combining  (\ref{equuz1}) and (\ref{equnnn}),  one
arrives at the well-known sum rule \cite{n1,quim,gronaur}
\begin{equation} \label{i}
{\rm I}) \;\; R_{0} - R + R_c -1 = k_1.
\end{equation}
There is, however, an interesting way of reading this sum rule, and it
is the following: if we take old data \cite{quim} and only central values 
to 
make the 
point clearer, sum rule (\ref{i}) will imply
\begin{eqnarray}  R_0=1+\underbrace{ n_+}_{
\mbox{\footnotesize + 0.21 }}=1+\underbrace{ u_{+}}_{
\mbox{\footnotesize  +0.00 }} -\underbrace{ z_{+}}_{
\mbox{\footnotesize  + 0.41 }} +{ k_{1}}  \nonumber
\end{eqnarray}
This means that in order for $k_1$, a quantity
of order $\epsilon^2$, could compensate for the values of $u_+$ and
$z_+$ (quantities of order $\epsilon$) and reproduce the
experimental value of $R_0=1+n_+$ ($R_0=1.21 \pm 0.35$ with old
data set), the value of $k_1$ (of ${\cal O}(\epsilon^2)$)  should
be $k_1 \sim 0.62$, which is indeed higher than the experimental
values of the other   quantities of order $\epsilon$ ($u_+ \sim 0.00$ and
$z_+\sim 0.41$). Of course, this is too naive, since we are taking only
central values and within errors (see Table \ref{expval}) everything is 
compatible.
However, this may  tell us that the central values of these
observables (CP-averaged branching ratios) are expected to change
a lot. Indeed if one repeats the experiment with the new data of CLEO
\cite{cleod}, BaBar \cite{babard,belled} and Belle \cite{belled}, again 
all 
experiments
are equally good, once errors are taken into account as they should.
However, it is interesting to notice that Belle's central values
seem to follow the sum rule wonderfully. If errors get reduced
with time we can start discriminating between the different
experiments.

\begin{table}[t]
\begin{center}
\begin{tabular}{|c|c|c|c|c|c|}
\hline Obs.& Order & Old & CLEO  & BaBar &
Belle\\
\hline $u_+=R-1$ & $\epsilon$ &
 $+.00\pm.18$ & $-.06\pm.28$ & $+.02\pm.15$ &
$+.16\pm.24$\\
$z_+=R_c-1$ & $\epsilon$ &
$+.41\pm.29$ & $+.27\pm.48$ & $+.27\pm.24$ &
$+.33\pm.37$\\
$n_+=R_0-1$ & $\epsilon$ &
$+.21\pm.35$ & $+.60\pm.79$ & $-.06\pm.37$ &
$-.18\pm.41$\\
$k_1$  & $\epsilon^2$ &
$+.62\pm.45$ & $+.93\pm.92$ & $+.19\pm.43$ &
$-.01\pm.53$\\
\hline
\end{tabular}
\caption{Sum rule parameters evaluated using old data \cite{quim} and 
CLEO, BaBar and Belle data \cite{cleod}--\cite{belled}. Notice that, for 
simplicity reasons, we are 
neglecting the small phase-space difference ($6\%$), in the sum rule 
expressions, between $B^{\pm} 
\rightarrow 
\pi^{\pm} K^0$ and $B_d^0 \rightarrow
\pi^{\pm} K^{\mp}, \pi^0 K^0$ affecting $R$ and $R_0$. 
}\label{expval}
\end{center}
\vspace*{-0.3truecm}
\end{table}
\medskip


Moreover, it is easy to understand the physical meaning of this
observable called $k_1$. From isospin one arrives at
\begin{eqnarray} \label{eqc}
-\sqrt{2} A \left( B^+ \rightarrow \pi^0 K^+ \right)&=& A \left(
B^+ \rightarrow \pi^+ K^0 \right) + d_1,
\nonumber \\
-A \left( B^0 \rightarrow \pi^- K^+ \right)&=&
A \left( B^+ \rightarrow \pi^+ K^0 \right) + d_2, \nonumber \\
\sqrt{2} A \left( B^0 \rightarrow \pi^0 K^0 \right)&=& A \left(
B^+
 \rightarrow \pi^+ K^0 \right)
 + d_2 - d_1,
\end{eqnarray}
where $d_i$ ($i=1,2$) are functions of $A_{\frac{1}{2} \;
1}=A_{\frac{3}{2} \; 1}$ and vanish if there is no isospin
breaking. In general one can write
\begin{equation} \label{di}
d_{i}=|P| \xi_i e^{i \theta_i } \left( e^{i \gamma} - a_i e^{i
\phi_{a_{i}} } -i b_i e^{i \phi_{b_{i}}} \right),
\end{equation}
where $P$ contain all CP-conserving terms of the  penguin
contribution to $B^+ \rightarrow \pi^+ K^0$. The $\xi_{i}$ parametrize
isospin breaking,  and they are expected to be small parameters.
$\theta_i$, $\phi_{a_{i}}$, $\phi_{b_{i}}$ are strong phases, and
$\gamma$ and $ib_{i}$ parametrize  weak phases  that change
 sign under a CP
transformation. We will follow  the notation
of \cite{neu3,quim}. We show explicitly in (\ref{di}) the dependence
on $\gamma$, meaning that $b_{1}$ and $b_{2}$ can be non-zero only
if there is new physics.

From the dependence of $k_1$ on the  $d_i$ parameters
\begin{eqnarray}
\label{defk1} k_1 &=& {2 \over x} \; \left( |{ d_{1}}|^2 +
|{\overline { d_{1}}}|^2 - {\rm Re} [{ d_{1}} \; {d_{2}^*}] - {\rm
Re} [{\overline { d_{1}}} \; {\overline { d_{2}}^{*}}] \right)
\end{eqnarray}
it is easy to interpret $k_1$ as a measure of the misalignment
 between the isospin-breaking
contributions to two channels: $\sqrt{2} A \left( B^+ \rightarrow
\pi^0 K^+ \right)$ and $A \left( B^0 \rightarrow \pi^- K^+
\right)$. Even in the presence of isospin-breaking if the new
contributions to these channels are equal, i.e. if  $d_1=d_2$, then
$k_1$ is exactly zero. On the contrary, if the isospin
contribution to these channels has opposite sign then $k_1$ 
is maximal. We should look at data to discern
 which of the two scenarios is closer to the one
realized in nature.
It is also possible to write down a completely general expression for 
$k_1$, valid for any model (see \cite{quim}).

\begin{figure}[t]
\vspace*{-1cm}
$$\hspace*{-1.cm}
\epsfysize=0.3\textheight \epsfxsize=0.3\textheight
\epsffile{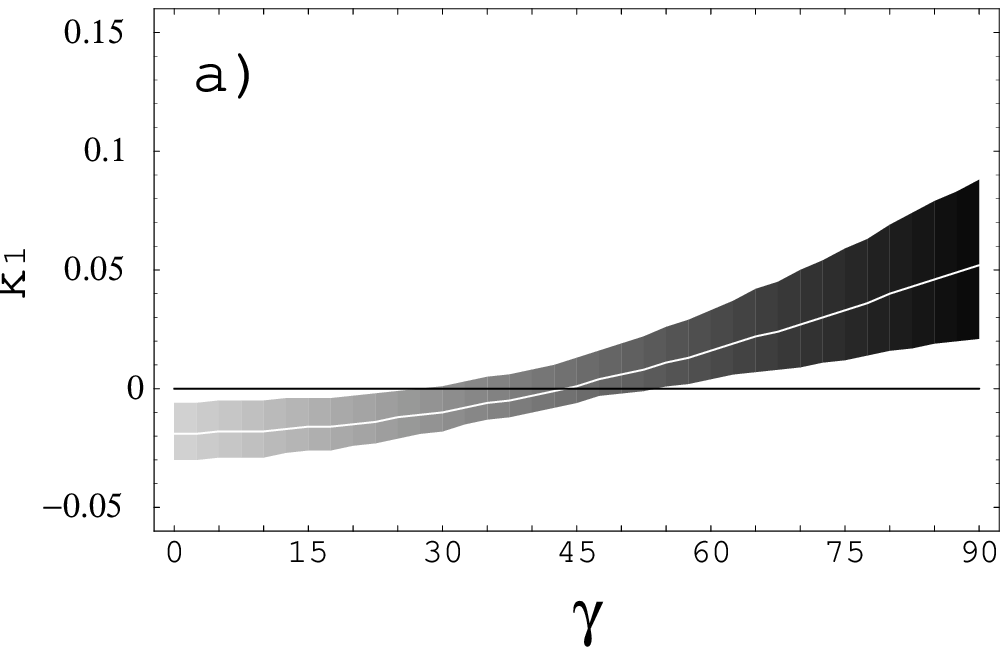} \hspace*{0.3cm}
\epsfysize=0.3\textheight \epsfxsize=0.3\textheight
 \epsffile{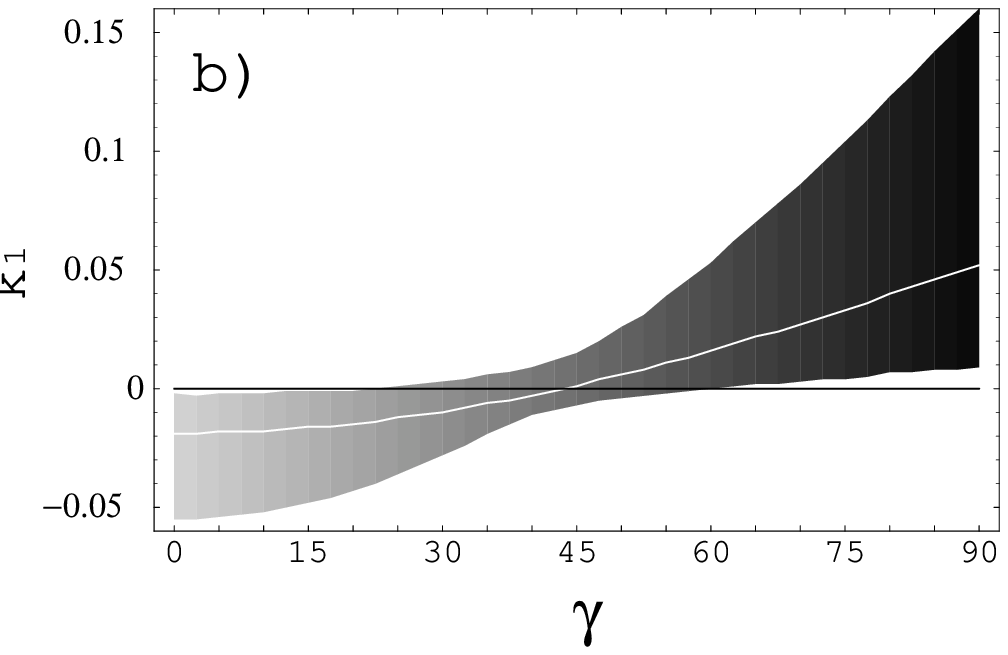}
$$
\vspace*{-0.9cm} \caption[]{Sum rule {\rm I} evaluated for the SM
using NLO QCD factorization \cite{mb} for values of $\gamma$
in the first quadrant: (a) low uncertainty ($\varrho_A=1$) from
annihilation topologies, (b) large uncertainty ($\varrho_A=2$)
from annihilation topologies} \label{k1plots}
\end{figure}

In order to have a reference value we show the prediction for
$k_1$ using NLO QCD factorization in the Standard Model. The two
plots of Fig.~\ref{k1plots} correspond to two different estimates of the 
uncertainty
coming from the annihilation topologies.

We can go beyond this sum rule and try to construct the simplest
sets of observables strongly correlated by isospin (we will number them 
III--V 
to follow the notation of \cite{quim}). They can help us 
in guessing  what we may expect from the data.
The result is the following sum rules \cite{quim}:
\begin{equation}
{\rm III}) \;\; R = {R_0 R_c}  +k_3,
\end{equation}
with $ k_3= z_+ \left(z_+-u_+\right)-k_1-k_1 z_+; $
\begin{equation}
{\rm IV}) \;\; R_c = -{R_0 \over R} + 2 +k_4,
\end{equation}
with $ k_4= {(u_+ z_+ + k_1) / (1 + u_+)}; $
and, finally,

\begin{table}[t]
\caption{Strongly correlated observables associated to sum rules
{\rm III--V}} \label{tab:tab1} \vspace*{0.1cm}
\begin{center}
\begin{tabular}{|lll|}\hline
\rule[-0.3cm]{0cm}{0.9cm} ${\rm III}$
 & ${\cal O}^{\rm III}_1=R$ & ${\cal O}^{\rm III}_2=R_0 R_c$  \\
\hline \rule[-0.3cm]{0cm}{0.9cm} ${\rm IV}$
& ${\cal O}^{\rm IV}_1=R_c$ & ${\cal O}^{\rm IV}_2=-R_0/R + 2$ \\
  \hline
\rule[-0.3cm]{0cm}{0.9cm} ${\rm V}$
 & ${\cal O}^{\rm V}_1=R_0$ & ${\cal O}^{\rm V}_2=-R_c/R + 2$ \\
 \hline
\end{tabular}
\end{center} 
\end{table}  
\begin{equation}
{\rm V}) \;\; R_0 = -{R_c \over R} + 2 +k_5,
\end{equation}
with $ k_5=k_1 + {u_+ (u_+ - z_+) /(1 + u_+)}$. This  sum rule can
be  related  with one proposed in~\cite{n1} (but with the inverse
$R/R_c$) for the SM case and in an approximate form, i.e. keeping
only the term $\xi^2_i$.
\begin{table}[t]
\begin{center}
\begin{tabular}{|c|c|c|c|}
\hline Observable &  CLEO  & BaBar &
Belle\\
\hline $k_3$ & $-1.10\pm1.31$  & $-0.17\pm0.51$
& $+0.07\pm0.60$\\
$k_4$ & $+0.97\pm0.91$ & $+0.19\pm0.43$ & $+0.04\pm0.51$\\
$k_5$ & $+0.95\pm0.90$ & $+0.18\pm0.43$ & $-0.03\pm0.50$ \\
\hline
\end{tabular}
\caption{Sum rules III--V evaluated using old data and
CLEO, BaBar and Belle data
}\label{tab:BPIK-obs}
\end{center}
\vspace*{-0.3truecm}
\end{table}
\medskip
\begin{figure}[t]
\vspace*{-.4cm}
$$\hspace*{-1.cm}
\epsfysize=0.3\textheight \epsfxsize=0.3\textheight
\epsffile{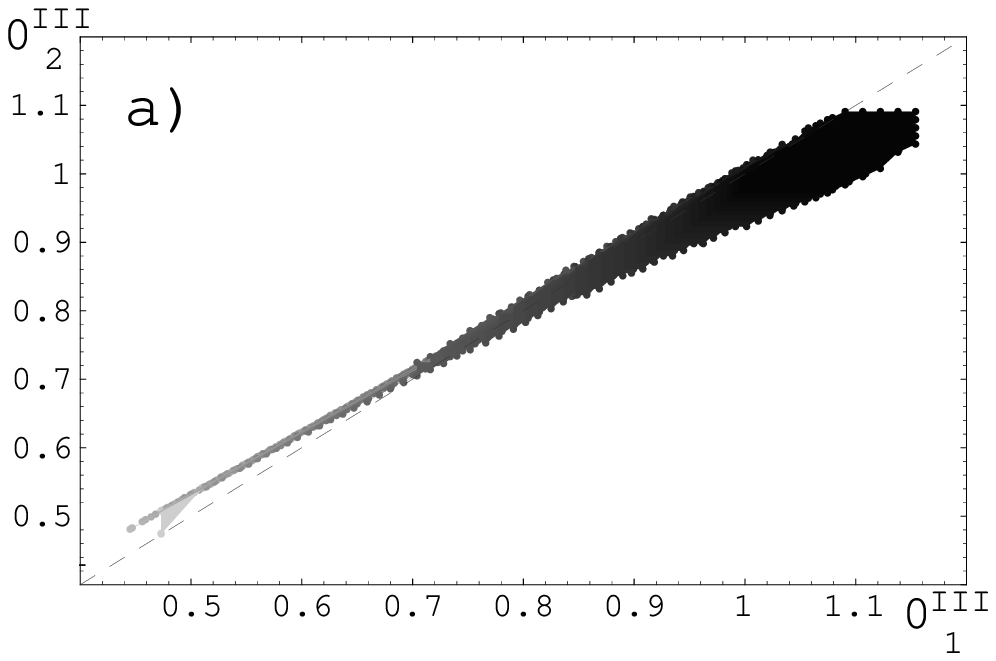} \hspace*{0.3cm} \epsfysize=0.3\textheight
\epsfxsize=0.3\textheight
 \epsffile{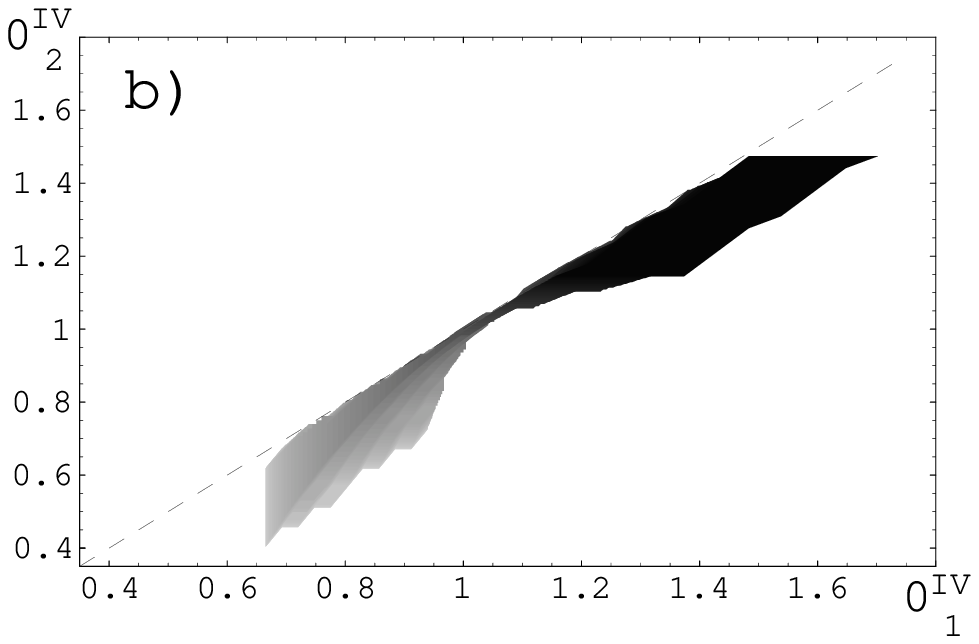}
$$
\vspace*{-0.9cm} \caption[]{Sum rules {\rm III} (a) and {\rm IV}(b)
evaluated for the SM using NLO QCD factorization \cite{mb} for
values of $\gamma$ in the first quadrant in the large uncertainty
case ($\varrho_A=2$) from annihilation topologies}\label{plotsobs}
\end{figure} 
%
%
%

The associated observables to these sum rules are given in Table 
\ref{tab:tab1}. The 
plots of associated observables (Fig.~\ref{plotsobs}) 
have an interesting interpretation:
in the absence of isospin-breaking,
 both observables should fall  in the diagonal of
Figs.~\ref{plotsobs}, with ${\cal O}_i^{\alpha}=1$.
If isospin breaking is small, $O_1^{\alpha}$ and $O_2^{\alpha}$
should stay near the diagonal. The deviation from 1
 {\bf along the diagonal} gives an idea of the
isospin-breaking terms of order $\xi_i$ (remember that $R$, $R_c$
and $R_0$
 measure isospin breaking of this
size). This is useful to have an idea of the maximal size of this
breaking. Notice that it also implies that each pair of
observables ($O_1^{\alpha}$, $O_2^{\alpha}$) is chosen in such a
way as to present the same deviation of order $\xi_i$, independently
of the model.

More interestingly, deviations {\bf from the diagonal}
 would
measure isospin-breaking
contributions of order  $\xi_i^2$.
It implies
that if the isospin  is not badly broken, we can estimate that the
 deviations from the diagonal will be smaller
than the square of the maximal deviation from 1 along the diagonal. For
 instance, in Fig.~\ref{plotsobs}b the
maximal deviation from 1 along the diagonal is approximately $0.5$;
the
maximal expected deviation from the
diagonal would then be $0.25$ and, indeed, this is the case.
This rule applies to all figures evaluated using NLO QCD factorization.
%
%
%


As in the case of $k_1$  the numerical values of the observables $k_3$, 
$k_4$ and $k_5$ are compatible, but their central values differ 
significantly (see Table \ref{tab:BPIK-obs}). Again Belle central values 
are 
in the expected ballpark. However experimental errors in these parameters 
are still too large to be conclusive.

\begin{table}[t]
\caption{Strongly correlated observables associated to sum rules
{\rm VI and VII}}
\label{tab:tab2}
\begin{center}
\begin{tabular}{|lll|}\hline
\rule[-0.3cm]{0cm}{0.9cm}
$\hspace*{-0.2cm}{\rm VI}$
& $\hspace*{-0.25cm}{\cal O}^{\rm VI}_1={\cal A}_{\rm CP}^{-+} R$ &
$\hspace*{-0.2cm}{\cal O}^{\rm
VI}_2=A_{\rm
CP}^{+0}-1+\left(1+ {\cal A}_{\rm CP}^{00} R_0-{\cal A}_{\rm CP}^{+0}
\right)\left(
1+ {\cal A}_{\rm CP}^{0+} R_c - {\cal A}_{\rm CP}^{+0} \right)$ \\
\hline
\rule[-0.3cm]{0cm}{0.9cm}
$\hspace*{-0.2cm}{\rm VII}$
& $\hspace*{-0.25cm}{\cal O}^{\rm VII}_1={\cal A}_{\rm CP}^{0+} R_c$ &
  $\hspace*{-0.2cm}{\cal
O}^{\rm VII}_2=A_{\rm
CP}^{+0}+\left({\cal A}_{\rm CP}^{-+} R-{\cal A}_{\rm CP}^{00} R_0
\right)/\left(
1+{\cal A}_{\rm CP}^{-+} R - {\cal A}_{\rm CP}^{+0} \right)$ \\
  \hline
\end{tabular}
\end{center}
\end{table}

In a similar way, one can also find a set of sum rules for the CP 
asymmetries, once the building blocks are identified: 
\begin{eqnarray}
{\cal A}^{-+}_{\rm CP} R
&=&{\cal A}^{+ 0}_{\rm CP} + u_{-},
\nonumber
\\
{\cal A}^{0 +}_{\rm CP}
R_c&=&{\cal
A}^{+ 0}_{\rm CP} + z_{-},
\nonumber
\\   
{\cal A}^{00}_{\rm CP}
R_0&=&{\cal
A}^{+ 0}_{\rm CP}+n_-,  \nonumber
\end{eqnarray}  
where the asymmetries are as defined in (\ref{eqabis}).
Using again isospin decomposition, we can demonstrate that:
\medskip
\begin{eqnarray}
u_- &\sim& {\cal O} (r) + {\cal O} (r^2, r\rho, q_C  
r)  \sim {\cal O}(\epsilon) \nonumber \\
z_- &\sim& {\cal O} (r_c) + {\cal O} (r_c^2, r_c\rho, q
r_c)  \sim {\cal O}(\epsilon) \nonumber 
\end{eqnarray}
and that  $n_- = u_- - z_- + k_2$ { with}
$k_2 \sim   
{\cal O}
(r_c r,...) \sim {\cal O}(\epsilon^2) $. From the expression of 
$n_-$ the sum rule follows automatically \cite{quim,n1}
$$ {\rm II) } 
{\cal A}^{00}_{\rm CP} R_0 -
{\cal A}^{-+}_{\rm CP} R+ {\cal A}^{0 +}_{\rm CP}
R_c
-{\cal A}^{+ 0}_{\rm CP} = k_2,$$
where $k_2$ can be related to the contributions to the different channels
\begin{eqnarray} \label{defk2}
k_2 = {2 \over x}\; \left(
|{ d_{1}}|^2 - |{\overline { d_{1}}}|^2 - {\rm Re} [{
d_{1}} \;
{d_{2}^*}]
+
{\rm Re} [{\overline { d_{1}}} \; {\overline { d_{2}}^{*}}]
\right) \nonumber
\end{eqnarray}
\begin{figure}[t] \vspace*{-1cm}
$$\hspace*{-1.cm}
\epsfysize=0.30\textheight
\epsfxsize=0.30\textheight
\epsffile{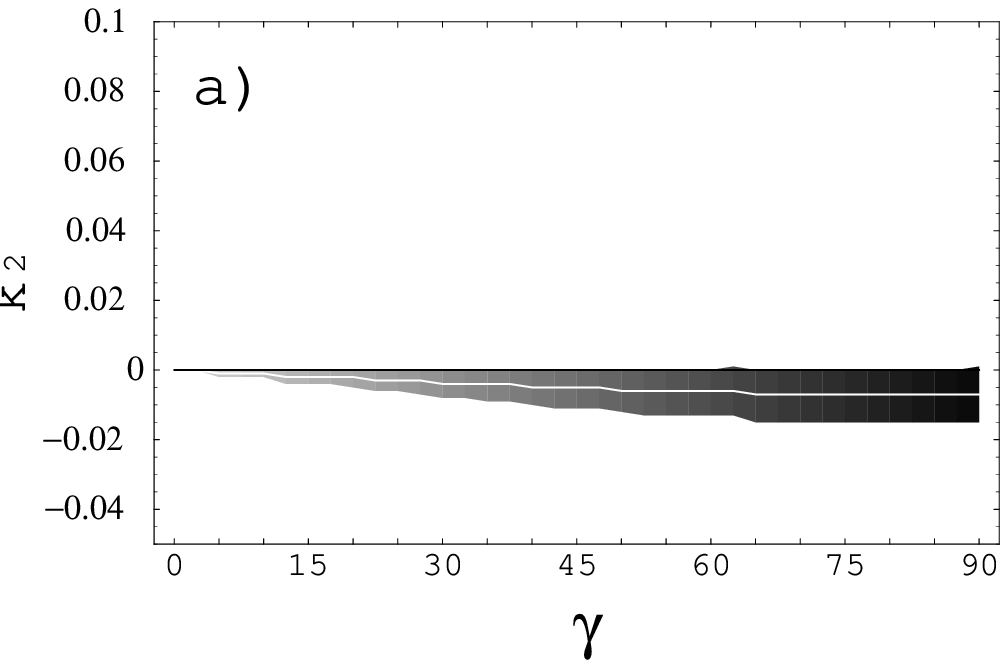} \hspace*{0.3cm}
\epsfysize=0.30\textheight
\epsfxsize=0.30\textheight
 \epsffile{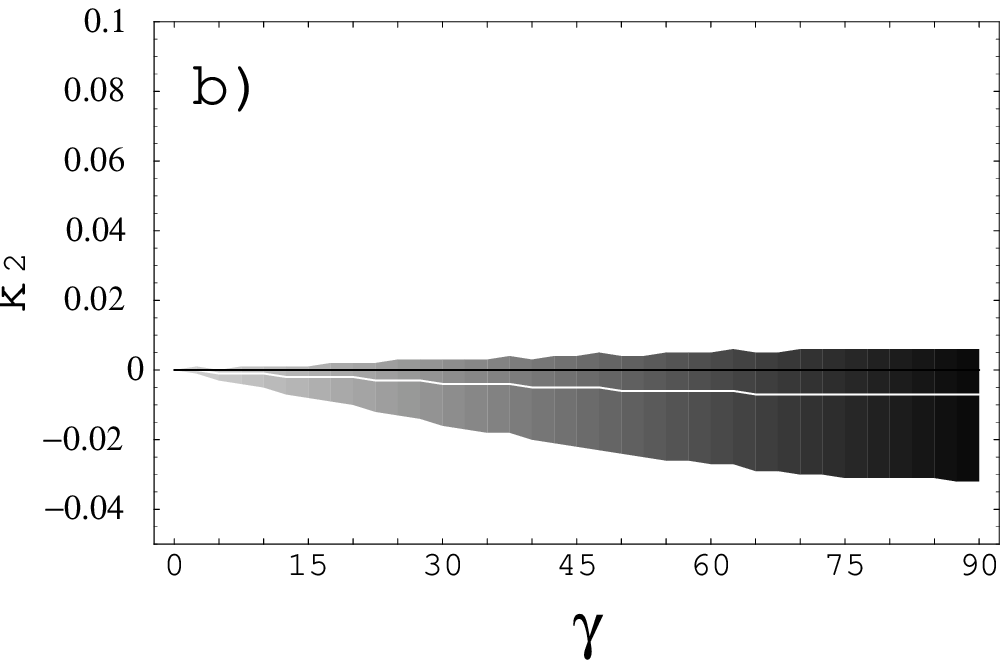}
$$
\caption[]{Sum rule {\rm II} evaluated for the SM  using
NLO QCD
factorization for
values of $\gamma$ in the first
quadrant: (a) low uncertainty ($\varrho_A=1$) from annihilation
topologies,
(b) large uncertainty ($\varrho_A=2$) from annihilation
topologies.
}\label{acp}
$$\hspace*{-1.cm}
\epsfysize=0.30\textheight
\epsfxsize=0.30\textheight
\epsffile{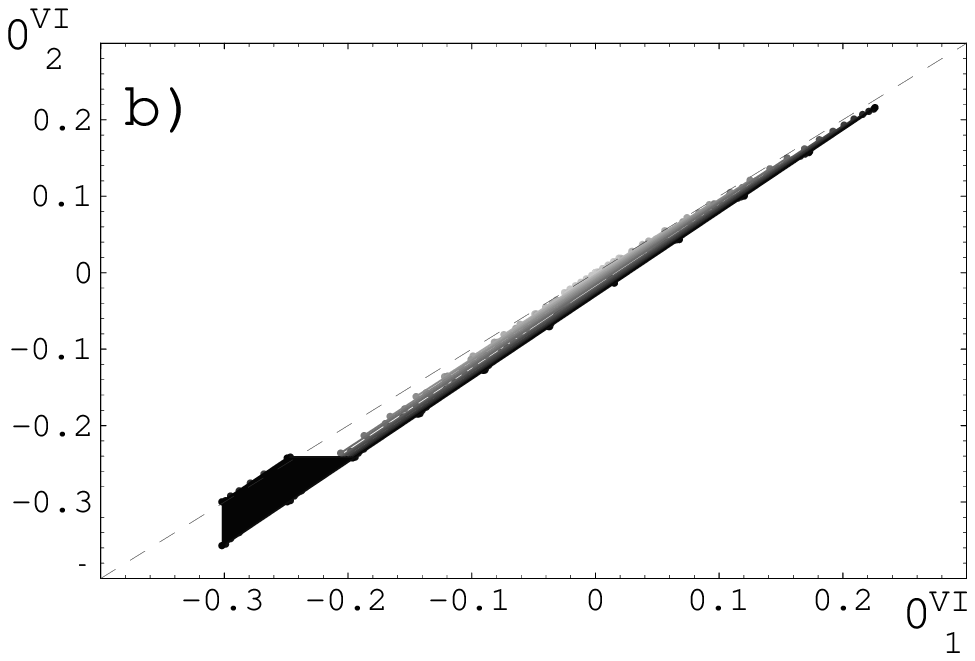} \hspace*{0.3cm}
\epsfysize=0.30\textheight
\epsfxsize=0.30\textheight
 \epsffile{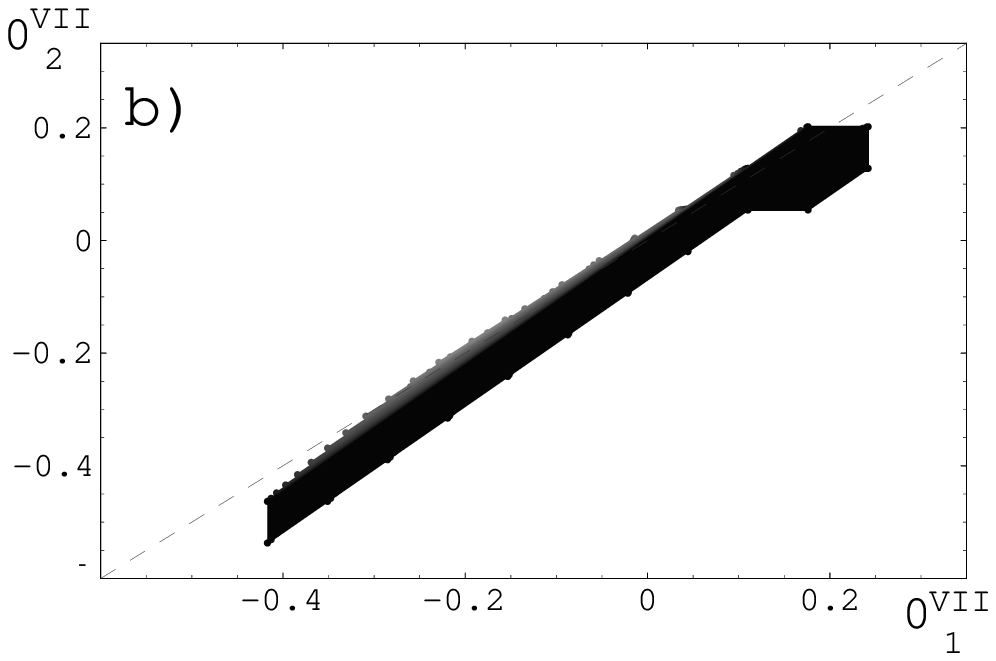}
$$
\vspace*{-1.4cm} 
\caption[]{Correlation between $O_1^{\rm VI,VII}$ and
$O_2^{\rm VI,VII}$.
} \label{correlacp}
\end{figure}
and it  also admits a nice interpretation: $k_2$ measures the
importance of weak
phase
differences between
$d_1$ and its CP conjugate, and between $d_2$ and its CP
conjugate. 
The same conditions that force $k_1$ to
vanish
also apply to $k_2$.
But, in addition, $k_2$ also vanishes
 if
$d_1={\overline d_1}$ and $d_2={\overline d_2}$.

We, also, show  as in the case of the CP-averaged branching ratios, the
prediction for the sum rule II evaluated using NLO QCD factorization in
Fig.~\ref{acp} and an example of strongly correlated observables using
CP asymmetries. In Table \ref{tab:tab2} these observables are defined and 
their prediction using QCD NLO factorization in the SM is illustrated in 
Fig.~\ref{correlacp} (see \cite{quim} for more details).

In conclusion we have shown that $B$-physics is becoming a powerful tool 
to
discriminate and even
exclude models. We have seen that  a certain type of left--right models 
with spontaneous 
CP
violation may be excluded by the new information obtained from $B$ 
physics. However, a last word about this specific model would 
still require, of 
course, to 
vary all input parameters (in particular CKM angles and quark 
masses).
On the hand, we have also shown that theory with a few 
reasonable hypotheses may help us in some
cases to `test'
data, for instance, in the  sum rules for $B\to \pi K$ decays. Still more
experimental precision is 
needed.

\section*{Acknowledgements} 
I acknowledge financial support by CICYT Research Project AEN99-0766 and 
FPA2002-00748, 
MCyT, 
Euridice HPRN-CT-2002-00311
and the 
Theory Division at CERN.




\end{document}